\documentclass[a4paper,twocolumn, nofootinbib,superscriptaddress,aps,prl,14pt,eqsecnum,notitlepage,showkeys]{revtex4-1}
\usepackage{multirow}
\usepackage{graphicx}
\usepackage{bm}
\usepackage{dcolumn}
\usepackage{hyperref}
\usepackage{gensymb}
\usepackage{amsmath}
\usepackage{dcolumn}    
\usepackage{ifpdf}
\usepackage{amssymb,lineno,amsfonts}
\usepackage{graphicx}   
\usepackage{bm}         
\usepackage{bbm}
\usepackage{mathrsfs}
\usepackage{upgreek}
\usepackage{mathtools}
\usepackage{epstopdf}
\usepackage{setspace}
\usepackage{hyperref}
\usepackage{natbib}
\usepackage{esvect}
\usepackage{amsmath}
\usepackage[usenames,dvipsnames]{xcolor}
\definecolor{med-blue}{RGB}{25,25,112}
\hypersetup{colorlinks, linkcolor={blue},citecolor={blue}, urlcolor={blue}}
\usepackage{times }
\usepackage [english]{babel}
\usepackage [autostyle, english = american]{csquotes}
\MakeOuterQuote{"}
\begin{document}
	\title{Temperature dependence of the anomalous Nernst effect in Ni$_{2}$MnGa shape  memory alloy}
	\author{Avirup De}
	\affiliation{Department of Physics, Indian Institute of Science Education and Research, Pune-411008, India}
	\author{Anupam K. Singh}
	\affiliation{School of Materials Science and Technology, Indian Institute of Technology (BHU), Varanasi-221005, India}
	\author{Sanjay Singh}
	\affiliation{School of Materials Science and Technology, Indian Institute of Technology (BHU), Varanasi-221005, India}
	\author{Sunil Nair}
	\affiliation{Department of Physics, Indian Institute of Science Education and Research, Pune-411008, India}
	\affiliation{Centre for Energy Science, Indian Institute of Science Education and Research,\\ Dr. Homi Bhabha Road, Pune, Maharashtra-411008, India}
	\date{\today}
	\begin{abstract}
	 We report a detailed investigation of the Ni$_{2}$MnGa shape memory alloy through magnetic, electronic, and thermal measurements. Our measurements of the anomalous Nernst effect (ANE) reveal that this technique is very sensitive to the onset of the pre-martensitic transition in sharp contrast to other transport measurements. With the ANE being sensitive to changes at the Fermi surface, we infer on the link between the structural modulations and the modulation of the Fermi surface via its nesting features, with the magnetic field induced strain being the mediating mechanism.             
	\end{abstract}        
	\pacs{Pacs}
	\maketitle
Ni$_{2}$MnGa shape memory alloy and its near stoichiometric compositions have been extensively investigated in the recent past due to their magnetic shape memory \cite{chmielus2009giant,ullakko1996large}, magneto-caloric \cite{PhysRevB.68.094401,qian2018tunable} and  magneto-resistive properties \cite{MR}. These systems, which are characterized by a complex interplay of structural, magnetic, and electronic degrees of freedom is also known to exhibit a large magnetic field induced strain (MFIS), in excess of $10\%$ \cite{strain,murray00}, which enhances its importance as a potential candidate for the development of efficient and fast magneto-mechanical actuators. The stoichiometric Ni$_2$MnGa has been widely used as a model system, and is understood to exhibit a well defined sequence of transitions below its Curie temperature  $T_C$ $\approx$ 376\,K \cite{webster1984magnetic,Brown_2002}.  At room temperature, Ni$_2$MnGa has a cubic $L2_1$ crystal structure ( $a= 5.825\AA$ and space group $Fm\bar{3}m$), which is referred to as the austenite phase \cite{Brown_2002}. Upon cooling, it transforms to a martensitic phase at  $T_M \approx 200$K via an intermediate pre-martensitic transition (PMT) which occurs at $T_{PM} \approx 260$K \cite{webster1984magnetic,Brown_2002}. In general, the pre-martensitic phase is considered to be a precursor state of the low temperature martensite phase with preserved cubic phase symmetry having a long period modulated structure\cite{Brown_2002,Singh_2013}.   The modulations originates from a periodic shuffling of the $(110)$ planes along the $[1\bar{1}0]$ direction with a periodicity of six atomic layers, also called the 3-fold or 3M modulation \cite{PhysRevB.54.15045,PhysRevB.92.054112,PhysRevB.51.11310}. On the other hand, the martensitic phase is characterized by a body-centered tetragonal \cite{webster1984magnetic,PhysRevLett.104.145702} or orthorhombic \cite{PhysRevB.90.014109,Brown_2002,PhysRevB.92.054112} crystal structure with much longer periodic modulation (5M \cite{PhysRevLett.104.145702,RIGHI20075237} or 7M \cite{PhysRevB.90.014109,Brown_2002,FUKUDA2009473}). The large magnetic field induced strain of Ni$_2$MnGa is closely linked with the modulation of the martensite phase\cite{strain,murray00} and the martensite phase appears through the modulated premartensite phase\cite{planes96,Singh_2013,PhysRevB.92.054112}, suggesting that the pre- martensitic phase is a perfect arena for the investigation of the possible cause and/or effects of these modulations \cite{PhysRevLett.119.227207}.
~However, this task is hampered by the fact that most of the conventional experimental tools are relatively insensitive across the PMT, thus limiting the ability to adequately characterize the structure-property relationships associated with these modulations. In this work, we report a detailed investigation of Ni$_{2}$MnGa system using a battery of magnetic, electronic, and thermal measurements. In contrast to other electronic and magnetic measurements, the anomalous Nernst effect (ANE) is seen to be sensitive to the PMT, bringing to the fore the potential utility of this technique in the investigation of such systems. 

\begin{figure}
	\centering
	\vspace{0cm}
	\hspace{0cm}
	\includegraphics[scale= 0.7]{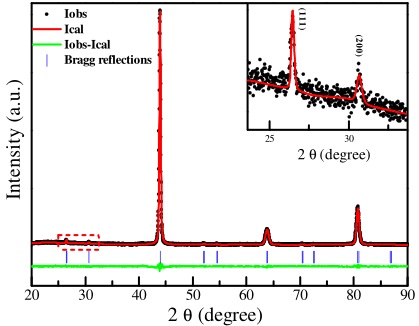}
	\caption{Rietveld profile fitting of the powder x-ray diffraction pattern of the Ni$_{2}$MnGa system. The inset shows an expanded view of the  low 2$\theta$ range, which confirms the $L2_{1}$ ordering.}
	\label{Fig1}
\end{figure} 

A polycrystalline ingot of Ni$_{2}$MnGa was prepared by the conventional arc-melting method using appropriate quantities of each constituent. The sample was melted several times to get uniform composition. The as-melted ingot was annealed in an evacuated quartz ampoule at a temperature of 1100\,K for 3 days to achieve further homogeneity and then quenched in an ice-water mixture. For compositional analysis, a small piece from the annealed ingot was evaluated using energy dispersive x-ray analysis. This data was collected at different regions of sample and the average composition was found to be Ni$_{1.96}$Mn$_{1.04}$Ga (hereafter termed as Ni$_{2}$MnGa). A part of the annealed bulk sample was ground into fine powder using a mortar pestle, sealed in evacuated quartz ampoule under Argon atmosphere, and annealed at 773\,K overnight to get rid of residual stresses introduced during the grinding \cite{singh2010struc, PhysRevB.92.020105, PhysRevB.91.134415}. This annealed powder was then used for x-ray diffraction (XRD) measurements. The small piece from annealed ingot was used for magnetization measurements, and the temperature dependent DC-magnetization data was collected using a vibrating sample magnetometer (Quantum Design, PPMS) in the field cooled cooling (FCC) and field cooled warming (FCW) protocols.

The Rietveld analysis of x-ray diffraction (XRD) data was carried using the FULLPROF software package and revealed that the system stabilized in a cubic structure (space group $Fm\bar{3}m$) . The Wyckoff positions  for Ni, Mn and Ga atoms used for refinements were 8a (0.25, 0.25, 0.25), 4a (0, 0, 0) and 4b (0.5, 0.5, 0.5), respectively. The experimental and calculated XRD plots are shown in Fig.1, which clearly reveal that the sample is in single phase, with all the Bragg peaks being well indexed. The inset shows the presence of (111) and (200) Bragg's reflections which is the evidence of $L2_{1}$ ordering. The refined lattice parameter is $a=5.8291(1) \AA$, which is in good agreement with the reported value \cite{Singh_2013} for Ni$_{2}$MnGa.

\begin{figure}
	\centering
	\vspace{0cm}
	\hspace{0cm}
	\includegraphics[scale= 0.6]{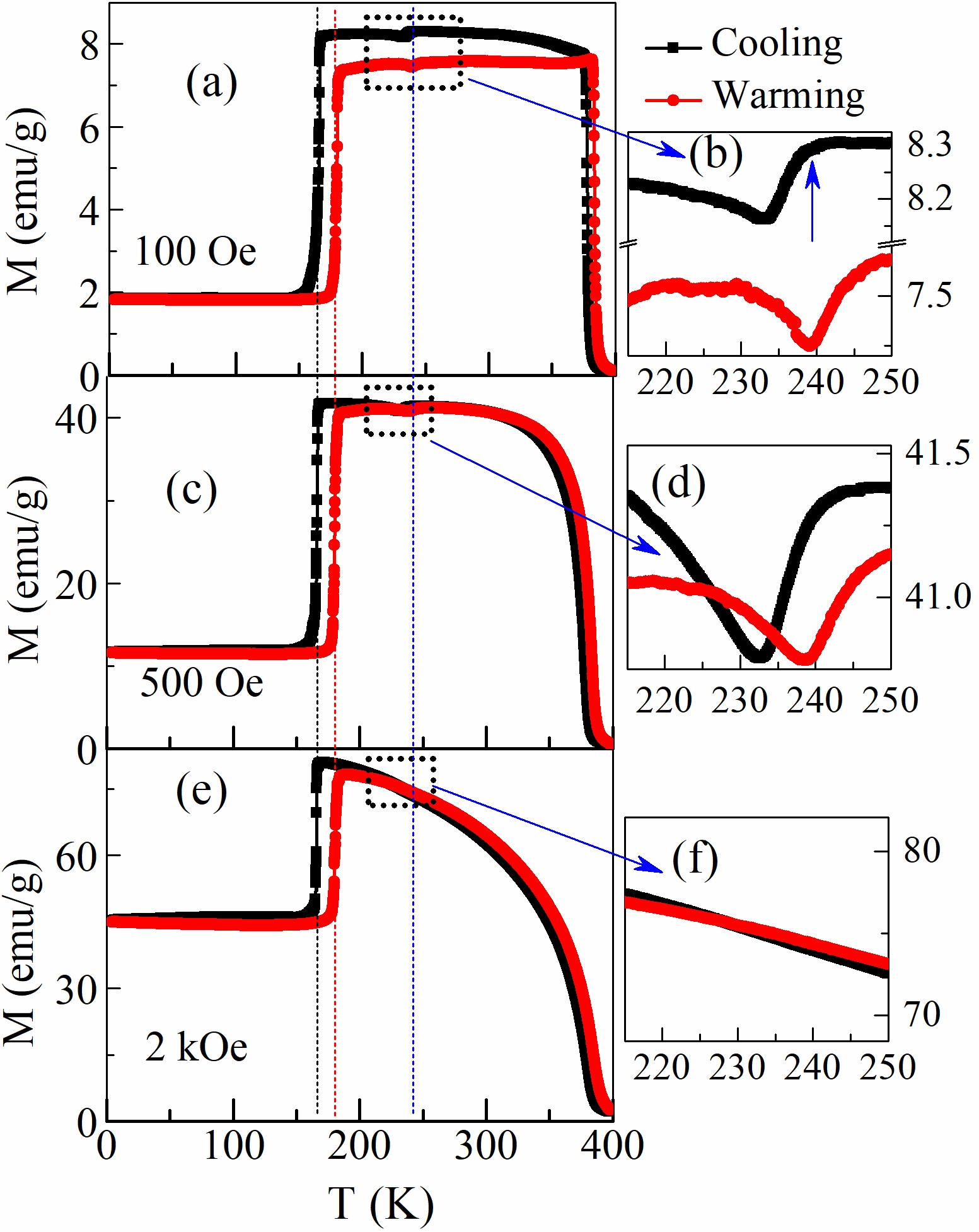}
	\caption{Temperature dependence of DC-magnetization at measuring fields of  (a) 100\,Oe,(c) 500\,Oe, and (e)2\,kOe $-$ with (b), (d), and (f) depicting a magnified part of the respective $M(T)$ around the pre-martensitic transition. The martensitic and pre-martensitic transitions are marked with dotted lines. }.
	\label{Fig2}
\end{figure}
 
Fig.2(a) illustrates the temperature dependent DC-magnetization of Ni$_{2}$MnGa as measured at low applied magnetic fields (100\,Oe), which confirms the sequential phase transitions in this alloy system as reported in literature \cite{PhysRevB.92.054112}. Upon cooling, the para-to-ferromagnetic, the austenite to pre-martensite and the pre-martensite to martensite transitions (MT) are observed at $383$K ($T_{C}$), $233$K ($T_{PM}$), and $167$K ($T_{M}$) respectively. During warming, all these transitions again appear in the reverse sequential order with a characteristic hysteresis at the martensite and pre-martensite phase transitions \cite{PhysRevB.80.144102, barandiaran2009magnetic, cui2004characteristics, wang2001effect,PhysRevB.92.054112}. In comparison with the previous reports on stoichiometric Ni$_2$MnGa \cite{webster1984magnetic,Brown_2002,magnetoelastic}, the $T_{PM}$ and the $T_{M}$ of the present composition is seen to be shifted to slightly lower temperatures, presumably due to a small variation in Ni and Mn content. These results are also in agreement with prior literature \cite{phasediagram1, phasediagram2}. Fig.2(c)\& 2(e) depict the temperature dependence of magnetization $M(T)$, measured at higher magnetic fields of 500\,Oe and 2\,kOe,respectively. Fig.2(c) shows that at a measuring field of 500\,Oe, the pre-martensite feature is slightly shifted to lower temperatures ( by around 1K) in comparison to the low field (100\,Oe) data. On the other hand, the pre-martensite feature is completely suppressed (Fig.2(e)) at a magnetic field value of 2\,kOe.  The suppression of this feature and the reduction of  $T_{PM}$ with increasing magnetic field is attributed to magnetoelastic coupling across the pre-martensite phase and is consistent with literature \cite{PhysRevLett.79.3926, magnetoelastic, cui2004characteristics,wang2001effect}. Fig.2(b), 2(d), and 2(f) depicts an expanded view of the magnetization-plots around the PMT (guided by blue arrows). The blue tick in Fig.2(a) shows the premartensitic start temperature ($T^{C}_{PMs}$ around 238K) in cooling. Apart from magnetic measurements, both the MT and the PMT are also observed in various other transport measurements, and the temperature dependence of resistivity ($\rho (T)$), thermal conductivity ($K(T)$), and the longitudinal Seebeck coefficient ($S_{xx}(T)$), measured in the absence of magnetic field, as depicted in Fig.3(a), 3(b), and 3(c) respectively. It is evident that these conventional transport measurements are relatively insensitive to the PMT. Nonetheless, the $\rho (T)$ curve confirms the metallic behavior of the sample. Similarly, the dominating contribution of the electron-type carriers to the longitudinal thermopower is also confirmed by the negative sign of the $S_{xx}(T)$ in the whole temperature range. 

\begin{figure}
	\centering
	\vspace{0.1cm}
	\hspace{-0.4cm}
	\includegraphics[scale=0.8]{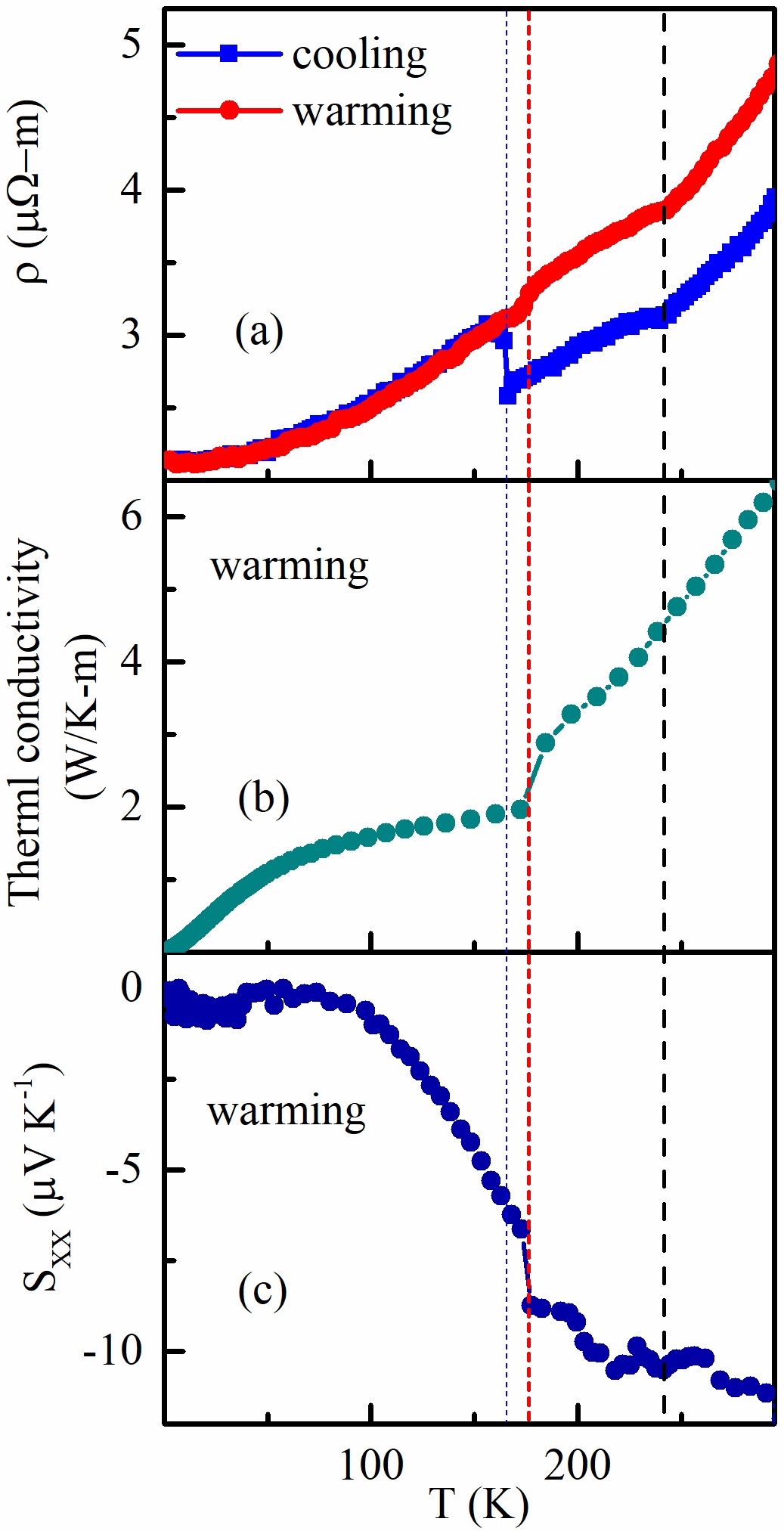}
	\caption{ Temperature dependence of (a) resistivity, (b) thermal conductivity, and (c) Seebeck coefficient as measured in the Ni$_{2}$MnGa system. The martensitic and pre-martensitic transitions are marked with dotted lines.}
	\label{fig3}
\end{figure} 

\begin{figure*}
	\centering
	\vspace{0.1cm}
	\hspace{-0.1cm}
	\includegraphics[scale=0.6]{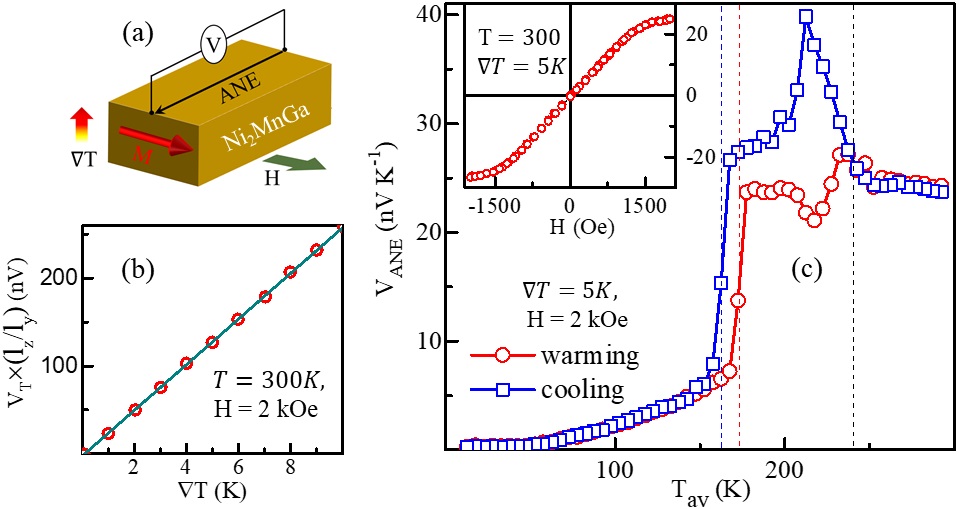}
	\caption{(a) depicts the device geometry of ANE. (b) illustrates the $\nabla T$ dependence of the ANE-signal. (c) depicts the temperature dependence of the ANE signal, $V_{ANE}(T)$  where the  $\nabla T$= 5\,K and $H = 2$\,kOe are kept fixed. In the inset of (c), the magnetic field dependence of $V_{ANE}$ signal at room temperature is shown, where the applied $\nabla T = 5$K. }
\end{figure*} 

The anomalous Nernst effect (ANE)  pertains to the generation of a transverse electric voltage ($V_T$) orthogonal to both the $\vec{M}$ and the applied thermal gradient ($\vec{\nabla}T$) across a magnetic metal $-$ where, the $\vec{\nabla}T$ acts as a driving force,  in analogy to an electric bias which causes the anomalous Hall effect \cite{PhysRevLett.101.117208,PhysRevB.96.174406}. It can be described by an empirical formula : $V_{T} = S_{xy} \vec{M} \times \vec{\nabla}T$, where $S_{xy}$ is defined as the transverse Seebeck coefficient \cite{PhysRevB.89.214406,PhysRevB.96.174406,doi:10.1063/1.5045262}. Fig.4(a) illustrates the geometry of our device, where the polycrystalline sample is cut and polished to form a rectangular shape of dimensions $7 \times 3 \times 1$ $mm^3$. The $\vec{\nabla}T$ is applied across the width ($l_z = 1$\,mm), the magnetic field is applied along the breadth ($l_x$), and the ANE voltage ($V_T$) is measured along the y-direction by using two gold wires at a distance of $3.6$ mm ($l_y$) between them. To cancel the background contributions, the magnetic field is reversed, and the ANE contribution is extracted as $V_T = \frac{1}{2}(V(+H)-V(-H))$. The average sample temperature is determined by $T_{av} = \frac{1}{2} (T_1+T_2)$ ; with $ \mid\vec{\nabla} T\mid = \nabla T = T_2-T_1$,  where $T1$ and $T_2$ are the temperatures at opposite surfaces of the sample, measured by two calibrated Cernox sensors. 

Fig.4(b) illustrates the linear dependence of $V_T$ on the applied $\nabla T$, where the sample temperature and the magnetic field are kept at 300\,K and 2\,kOe, respectively. The linearity of the $V(T)$ vs. $\nabla T$-plot along with a zero intercept confirms the absence of any thermal loss in our measurement. Fig.4(c) presents the temperature dependence of the normalized ANE signal so that $V_{ANE}(T) = (\frac{V_{T}}{\nabla T} )(\frac{l_z}{l_y})$, where the applied magnetic field ($\vec{H}$) and the applied thermal gradient ($\vec{\nabla} T$) are kept fixed at 2\,kOe and 5\,K, respectively. In the inset of Fig.4(c), the magnetic field dependence of the $V_{ANE}$ is depicted, where $T_{av}$=300\,K and $\nabla T$=5\,K are kept fixed. Interestingly, a reasonably large ANE signal is observed to appear at room temperature itself. In agreement with the previous reports on Heusler alloys \cite{doi:10.1063/1.5045262}, the present Ni-Mn-Ga system also shows a significant drop in the $V_{ANE}(T)$-signal across the MT. Moreover, the $V_{ANE}(T)$ turns out to be very sensitive across the PMT, although such anomaly in the dc magnetization is substantially suppressed across the PMT at a field of 2\,kOe. 

The origin of ANE is well understood in the picture of an effective spin-orbit interaction $-$ which is further decomposed into three possible ingredients: namely the Berry curvature, the side jump, and the skew scattering mechanisms \cite{PhysRevB.96.174406,PhysRevLett.101.117208}. For our analysis, we consider Mott's relation that connects the transverse thermoelectric coefficient, $\alpha_{xy}$, to the transverse electrical conductivity, $\sigma_{xy}$ through the relation: $\alpha_{xy} = \frac{\pi^2 k_B^2 T}{3e} \times (\frac{\partial \sigma_{xy}}{\partial E})_{E_F}$, where $k_B$, $e$, $E$, and $E_F$ are the Boltzmann constant, electric charge, energy, and the Fermi energy, respectively \cite{PhysRevB.96.174406,PhysRevB.89.214406,PhysRevLett.101.117208,doi:10.1063/1.5045262}. This implies that $\alpha_{xy}$ depends on both the temperature and the slope of the $\sigma_{xy}$ tensor with the Fermi-surface, or $(\frac{\partial \sigma_{xy}}{\partial E})_{E_F}$. Therefore, the $S_{xy}$, being linked to $\alpha_{xy}$ through the relation: $S_{xy} = \rho (\alpha_{xy} - \sigma_{xy}S_{xx})$ \cite{doi:10.1063/1.5045262,PhysRevB.96.174406,PhysRevB.89.214406}, also becomes sensitive to the changes in the Fermi surface. Interestingly, a previous report has also attested to the sensitivity of the Nernst signal on the Fermi-surface distortions \cite{sensitiveprobe}.  

The large MFIS of the Ni$_{2}$MnGa is intimately linked with the existence of a long period modulated structure. However, the origin of those modulations is still under debate $-$ and two different scenarios have been suggested. In the Adaptive model, the modulation is portrayed as an effective twinning emerging from the minimization of elastic energy \cite{PhysRevLett.104.145702,PhysRevB.43.10832}. An alternative mechanism is the soft-phonon mode-based displacive modulation model, which is further linked to the Fermi surface nesting features. The latter has received more theoretical and experimental support \cite{PhysRevB.54.15045,PhysRevB.51.11310,PhysRevB.64.024305,Shapiro_2007,CDW,PhysRevB.92.054112,PhysRevB.91.134415,PhysRevLett.100.165703,PhysRevB.68.134104,PhysRevB.66.054424}. 

Inelastic neutron scattering experiments have confirmed the signature of phonon softening \cite{PhysRevB.54.15045,PhysRevB.51.11310,PhysRevB.64.024305} and the phasons \cite{Shapiro_2007,PhysRevB.68.134104} associated with the CDW resulting from the Fermi surface nesting at both the pre-martensite and the martensite phases. Recently, photoemission spectroscopy has also inferred on the occurrence of CDW at the onset of the PMT \cite{CDW}. This partial nesting of the Fermi surface at the pre-martensite phase has also been confirmed by electron-positron annihilation experiments \cite{Haynes_2012}.  High resolution X-ray diffraction studies have also revealed non uniform atomic displacements in the modulated phase, phason broadening in the satellite peaks, and a temperature variation of the modulation wave vector without any commensurate lock in phase \cite{PhysRevB.90.014109, PhysRevB.92.054112}. Thus, it can be concluded that modulation in pre-martensite and martensite phase for Ni$_{2}$MnGa is in all probability linked with the soft phonon based model associated with a Fermi surface nesting. Our results which depict a remarkable sensitivity of the ANE signal to the PMT is presumably due to a change in the Fermi surface via its nesting features, which occurs at the onset of the PMT. It is also important to note that the possibility that  a contributory factor could be a variation of  $\sigma_{xy}$ $-$ due to its crucial connection to $S_{xy}$. The $\sigma_{xy}$ could also be expressed in terms of transverse resistivity ($\rho_{xy}$) via the relation $\sigma_{xy}=\frac{-\rho_{xy}}{(\rho^2+\rho^2_{xy})} \approx -\frac{\rho_{xy}}{\rho^2}$ \cite{PhysRevLett.101.117208}, where $\rho$ is the longitudinal resistivity ($\rho \equiv \rho_{xx}$). Thus, with the scaling relation $\rho_{xy} \propto M \rho^n$, where n is a number that typically varies within $0$ to $2$ \cite{PhysRevLett.101.117208,PhysRevB.96.174406,nair2012hall,PhysRevLett.99.086602} $-$ it could be shown that the role of $\sigma_{xy}$ in the large anomaly in $V_{ANE}(T)$ across the PMT would be negligible since the variation of both $M(T)$ \& $\rho (T)$ are faint across the PMT. Thus, the signatures in the measured ANE are more likely to reflect a change in the Fermi surface that occurs at the onset of PMT. It is to be noted that this nesting feature does also exist in the martensite phase \cite{Haynes_2012,PhysRevB.68.134104,PhysRevB.66.054424}. However, its contribution to the ANE-signal cannot be estimated due to the overwhelming change in the ANE arising as a consequence of the structural transformation ( and associated change in the magnetization) at the MT.  

Apart from $T_{PM}$ and $T_M$, the $V_{ANE}(T)$ also hints to the presence of another characteristic temperature $T^*$ $-$ well below the onset of the PMT $-$ across which a significant change in the ANE voltage is observed.  We note that in the cooling data (Fig.4(c)),  V$_{ANE}(T)$ starts to rise at around 238K  -  the starting point of pre-martensite phase transition ($T^{C}_{PMs}$) as is also observed in our low field magnetization data (Fig.2(a)). The ANE-voltage continues to rise until it is fully in the pre-martensite phase at $T^*$.  Below this temperature,  $V_{ANE}(T)$  starts to decrease - presumably due to local traces of the martensite phase being present together with the pre-martensite phase just below of $T^*$. Previous NMR experiments have also inferred on such a temperature regime between the $T_{M}$ and $T_{PM}$, where both the pre-martensitic and the martensitic phases coexist \cite{PhysRevB.91.134415}. On, further cooling, the main martensitic transition occurs, which is also observed in all the measurements, characterized by $T_{M}$, with a reasonable thermal hysteresis in the $V_{ANE}(T)$ being observed across both the MT and the PMT.  
     
Though the Mott relation can explain the ANE signal, it also suggests that $S_{xx} \propto (\frac{\partial \sigma_{xx}}{\partial E})_{E_F}$ \cite{PhysRevLett.101.117208,doi:10.1063/1.5045262,PhysRevB.96.174406}. This implies that  $S_{xx}$ should be equally sensitive to the onset of the PMT, whereas this is clearly not the case, as is evident from Fig.3(c). We speculate that this discrepancy could be a consequence of the externally applied magnetic field in the ANE measurements. Measurements of $S_{xx}(T)$, are performed in the absence of an external magnetic field, where a sizable electronic background of the unchanged Fermi surface dominates over that arising from the weak nesting at this pre-martensitic phase. On the other hand, as a consequence of the measurement protocol, the ANE is more sensitive to changes \emph{due} to the magnetic field.  Interestingly, at the same measuring field of 2\,kOe,  the dc magnetization data did not exhibit  a significant feature across the PMT. Here, we note that the ANE has been reported to exhibit features that are beyond the magnetization scaling in some single crystalline specimens, with these features being linked with the role of the effective Berry curvature \cite{guin2019anomalous, sakai2018giant}. As our sample is a polycrystalline slab, the contribution from the Berry curvature mechanism in the measured ANE signal is not likely to be significant. We speculate that the observed features in the vicinity of the PMT arises due to the coupling between the magnetism and the Fermi surface, with the magnetism being tuned by magnetic field driven changes in the Fermi surface. Recently, G. Lantz and co-workers also have inferred about such a coupling in their pump-probe experiments on  stoichiometric Ni$_2$MnGa specimens \cite{PhysRevLett.119.227207}. Their observations suggested that photo-induced demagnetization modified the Fermi surface, as a consequence of which the nesting vector and the modulation periodicity were also tailored.
 
In conclusion, we present a detailed investigation of Ni$_{2}$MnGa  using a battery of magnetic, electronic and thermal measurements. Across the PMT, a predominant suppression of the anomaly in the $M(T)$ curve is observed at higher magnetic fields $-$ suggesting the existence of magnetoelastic effect across the PMT. The ANE is presented for the first time in this Ni-Mn-Ga class of materials, and it clearly shows a significant  sensitivity across the PMT. This is in sharp contrast to other conventional transport measurements, like $\rho (T)$, $K(T)$, and $S_{xx}(T)$, where the PMT is barely discernible. With the ANE being sensitive to changes in the Fermi surface, we infer that such sensitivity of the $V_{ANE}(T)$ is due to the changes in the Fermi surface $-$ which further support the soft phonon based model. Finally, the measured $V_{ANE}(T)$ in conjunction with  both the $M(T)$ and the $S_{xx}(T)$ data also supports the possibility of coupling between magnetism and the Fermi surface, which could have significant implications in understanding these functional materials. Our results also suggest that the ANE could be utilized as a powerful tool for more detailed investigations in the Ni-Mn-Ga and related systems with different compositions and/or forms, like, single crystals or thin films.  
  
AD is thankful to Shruti Chakravarty for her support.  AD acknowledges UGC, Govt. of India for providing financial support through Senior Research Fellowship (SRF). The authors acknowledge funding support by the Department of Science and Technology (DST, Govt. of India) under the DST Nanomission Thematic Unit Program (SR/NM/TP-13/2016).  S.S.\ thanks Science and Engineering Research Board of India for financial support through an Early Career Research Award (Grant No.\ ECR/2017/003186) and the award of a Ramanujan Fellowship (Grant No.\ SB/S2/RJN-015/2017). 

\bibliography{Bibliography}
 
\end{document}